\documentclass[12pt,epsfig]{article}
\usepackage{graphicx}

\usepackage[latin1]{inputenc}
\usepackage{amsmath}
\usepackage{amsfonts}
\usepackage{amssymb}
\usepackage{graphicx}
\DeclareGraphicsExtensions{.pdf,.png,.jpg}

\usepackage{epstopdf}
\def\beq{\begin{equation}}
\def\eeq{\end{equation}}
\def\bea{\begin{eqnarray}}
\def\eea{\end{eqnarray}}

\begin{document}

\begin{center}
	{\Large \bf Self-similar features around spectral singularity  in complex barrier potential
	}

\vspace{1.3cm}

{\sf   Mohammad Hasan  \footnote{e-mail address: \ \ mhasan@isac.gov.in, \ \ mohammadhasan786@gmail.com}$^{,3}$,
 Bhabani Prasad Mandal \footnote{e-mail address:
\ \ bhabani.mandal@gmail.com, \ \ bhabani@bhu.ac.in  }}

\bigskip

{\em $^{1}$Indian Space Research Organization,
Bangalore-560213, INDIA \\
$^{2,3}$Department of Physics, Institute of Science,
Banaras Hindu University,
Varanasi-221005, INDIA. \\ }

\bigskip
\bigskip

\noindent {\bf Abstract}
\end{center}
The spectral singularity (SS) from a non-Hermitian potential is one of the most remarkable scattering feature of non-Hermitian quantum mechanics. At the spectral singular point, the scattering amplitudes diverge to infinite. This  phenomena have been extensively studied  over the last two decades. The previous studies have suggested the need of extremely fine control of the various system parameters for practical applications of SS. However no study have been carried out to understand the behavior of the scattering amplitude in the vicinity of SS. Such study would be important towards the practical application of SS where it is desired to maintain outgoing scattering amplitude to a specific value. The behavior of the loci of constant scattering amplitude in the neighborhood of SS are studied for a pure imaginary barrier potential $iV$, $V \in R^{+}$. We show that these loci are elliptical and self-similar to each other in $E-V$ plane where $E$ is energy of the wave. The orientations of these ellipses for reflection ($R$) and transmission ($T$) amplitude around a given SS are same. This is surprising due to the different mathematical forms of $R$ and $T$. In the vicinity of SS, $R \approx T$.

\medskip
\vspace{1in}

\newpage

\section{Introduction}
The study of  non-Hermitian systems in quantum mechanics started as a mathematical curiosity. Soon it was realized that a fully consistent quantum theory can be developed for non-Hermitian system by restoring the equivalent Hermiticity and maintaining the unitarity of time evolution in a modified Hilbert  space \cite{ben4}-\cite{benr}. Due to this the non-Hermitian quantum mechanics (NHQM) has become a topic of frontier research in the last two decades.    Due to analogy of the Schrodinger equation and certain wave equation in optics, the phenomena of non-Hermitian quantum mechanics (NHQM) can be mapped to the analogous phenomena in optics. This has lead to the possibility of experimental observation of the predictions of NHQM. This has been indeed the case and the predictions of NHQM have been observed in optics   \cite{ opt1}-\cite{kotto}. The experimental realization of non-Hermitian system have further ignited huge interest to study the subject both theoretically and experimentally. The study of non-Hermitian system in optics has become a constant theme of further research and advancement in the subject. 

One of the most beautiful finding of NHQM is the discovery of coherent perfect absorption (CPA)  also known as anti-laser. In CPA, two coherent beam of light interfere each other in such a way as to completely cancel each other out within the non-Hermitian medium. This enable a complete absorption of coherent light of specific frequency for a given system. There is an interplay in the absorption and the interference effects for the occurrence of CPA. CPA has also been demonstrated experimentally. For the first time it was demonstrated in a silicon slab \cite{cpa_first} and later with the use of metamaterial \cite{cpa_metamaterial}. Very recently CPA has also been shown for disordered medium \cite{cpa_disordered}. For Hermitian system CPA is absent due to unitary nature of the interaction.     

CPA is the time-reversed partner of spectral singularity (SS) which is a new type of resonance pertaining to the non-Hermitian system. At SS energy, the determinant of the scattering matrix vanishes leading to the blow up of the reflection and transmission amplitude . It was noted that the mathematical concept of this singularity has physical realization as special type of zero-width resonance for non-Hermitian scattering potential \cite{ali_09}. This led to a flurry of investigations by many authors \cite {ss_investigation1} -\cite{nhsfqm}. The concept of CPA being the time-reversed partner of SS has formed the mathematical basis of the investigation of CPA action from a medium \cite {longhi_2010_a, longhi_2010_b}. CPA has been realized in various experimental setups \cite{cpa_setup1, cpa_setup2, cpa_setup3}. 

It has been noted that the occurrence of SS (or CPA) are extremely sensitive to the system parameters such as gain, loss parameters and the spatial extent of the system \cite {ali_09,ali_11_spherical_gain}. Due to technical limitations of such a system, the physical parameters of the system may not be exactly tuned to the value which precisely give rise to SS (or CPA).  The system parameters are expected to fluctuate around the value that support SS. Therefore for the purpose of practical applications , it is important to understand the behavior of reflection and transmission amplitude around the spectral singular point with respect to small change of system parameters. Moreover for the usefulness of a device that display CPA or SS, it is desirable to keep the scattering amplitude constant. The drastic reduction of scattering amplitude has been noted in \cite {ali_09} when system parameters infinitesimally changes from those values that supports SS.

In this paper we  study  the variation of tunnelling ($T$)and reflection ($R$) amplitude  around SS from a non-Hermitian system that support SS. We consider complex barrier potential `$+iV$', $V \in R^{+}$ of finite width $b$ and study the two dimensional variation of $T$ and $R$ in the neighborhood of $(E_{ss}, V_{ss})$ in $E-V$ plane where $E$ is the energy of the incident plane wave. Here $E_{ss}$ is the spectral singular energy for potential $V=i V_{ss}$, $V_{ss} >0$. The problem is mathematically complicate even for this simplest (finite extent) non-Hermitian system and we perform analytical calculations wherever possible and revert to numerical calculations when it is not possible to advance analytically due to the transcendental nature of the problem. It is found that for a given $(E_{ss}, V_{ss})$, both the constant-$T$ and constant -$R$ loci in the immediate neighborhood of $(E_{ss}, V_{ss})$ are ellipses. It is further noted that all such ellipses for different values of $T$ and $R$ around the vicinity of SS are self-similar. The orientations of these ellipses for both $T$ and $R$ are same which is a very surprising behavior considering the different mathematical forms of $T$ and $R$ as well as the analytical equations of both the loci of $T$ and $R$. Further $T \approx R$ in the vicinity of SS i.e  the loci of constant-$T$ and constant -$R$ overlap in the vicinity of SS.

We organize our paper as follows. In section \ref{ss_inro} we introduce the reader about spectral singularity (SS). In section \ref{ss_barrier} we discuss SS from a complex barrier potential.  Section \ref{ss_near} is devoted to our detailed analysis of the behavior of transmission ($T$)and reflection ($R$) amplitude in the neighborhood of SS for a complex barrier potential `$+iV$', $V \in R^{+}$. In this section we demonstrate the self-similar behavior of $T$ and $R$  around SS. We discuss the results in section \ref{result_discussion}.

\section{Spectral singularity}
\label{ss_inro}
The Hamiltonian operator in one dimension for a non-relativistic particle is (in the unit $\hbar=1$ and $2m=1$)
\begin{equation}
H=-\frac{d^{2}}{d x^{2}}+ V(x)
\end{equation}
where $V(x)=V_{1}(x)+ i V_{2}(x)$ and ${V_{1}, V_{2}} \in R$. $V_{1}(x), V_{2}(x) \rightarrow 0$ as $x \rightarrow \pm \infty$. If $\int U (x) dx$, where $U(x)=(1+\vert x \vert) V(x)$ is finite over all $x$, then the Hamiltonian given above admits a scattering solution with the following asymptotic values
\begin{eqnarray}
\psi (k,x \rightarrow +\infty)= A_{+}(k) e^{ikx}+B_{+}(k) e^{-ikx} \\
\psi (k,x \rightarrow -\infty)= A_{-}(k) e^{ikx}+B_{-}(k) e^{-ikx}
\end{eqnarray}    
In the above $k=\sqrt{E}$. The coefficients $A_{\pm}, B_{\pm}$ are connected through a $2 \times  2$ matrix $M$, called as transfer matrix.
\beq
\begin{pmatrix}   A_{+}(k) \\ B_{+}(k)     \end{pmatrix}= M(k) \begin{pmatrix}   A_{-}(k) \\ B_{-}(k)    \end{pmatrix} 
\eeq
where,
\beq
 M(k)= \begin{pmatrix}   M_{11}(k) & M_{12}(k) \\ M_{21}(k) & M_{22}(k)   \end{pmatrix}  
\eeq
The transmission and reflection coefficients are obtained in terms of the elements of transfer matrix as given below,
\begin{equation}
t_{l}(k)=\frac{1}{M_{22}(k)} \ \ , r_{l}(k)=-\frac{M_{21}(k)}{M_{22}(k)}
\label{tl_rl}
\end{equation} 

\begin{equation}
t_{r}(k)=\frac{1}{M_{22}(k)}=t_{l}(k) \ \ , r_{r}=\frac{M_{12}(k)}{M_{22}(k)}
\label{tr_rr}
\end{equation}
From above equations it is seen that the reflection and transmission amplitudes will simultaneously blow up to infinite at real $k=k_{ss}=\sqrt{E_{ss}}$ if $k_{ss}$ is a real zeros of $M_{22}(k)=0$. The simultaneous divergence of reflection and transmission amplitudes are referred as `spectral singularity' and is associated with zero width resonance of eigenvalue equation $H \psi =k^{2} \psi$. At $E=E_{ss}$, both $R$ and $T$ are infinite.

\section{Spectral singularity of complex barrier potential}
\label{ss_barrier}
\begin{figure} 
\begin{center}
\includegraphics[scale=0.25]{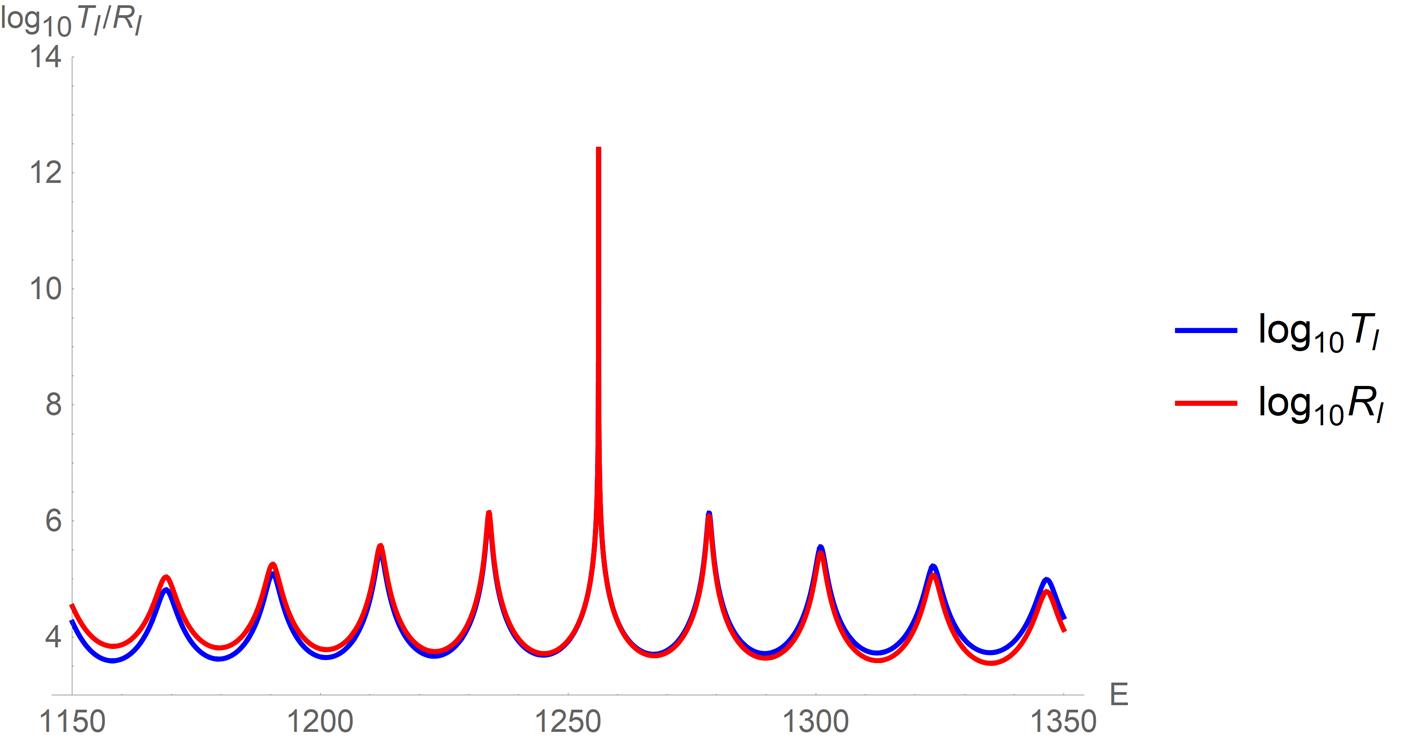} 
\caption{\it The spectral singularity for the potential $V=6.15055+35.000000045i$ of width $b=10$ at energy around $E=1256.158448057$. For $E> E_{ss}$. Both $R$ and $T$ diverges at spectral singularity.} 
\label{fig_ss_complex_barrier}
\end{center}
\end{figure}

For a non-Hermitian barrier potential  $V(x)=V=V_{1}+iV_{2}$, $\{V_1,V_2\} \in R$ over the interval $(0,b)$ and zero elsewhere, the transfer matrix has the following elements 
\begin{eqnarray}
M_{11}(b)  &= &  (\cos{ k' b} +u_{+}\sin{ k'b})e^{-ik b}  \label{m11_barrier} \\
M_{22} (b) &=& (\cos{ k' b} - u_{+}\sin{ k'b})e^{ik b} =M_{11}(-b) \label{m22_barrier} 
\end{eqnarray}
and the off-diagonal elements
\begin{eqnarray}
M_{12}(b) &=& u_{-}\sin{ (k'b)} e^{-ik b}    \label{m12_barrier} \\
M_{21}(b) &=& -u_{-}\sin{ (k'b)} e^{ik b} =M_{12}(-b) \label{m21_barrier}
\end{eqnarray}
where $k'=\sqrt{E-V}$ and,
\begin{eqnarray}
u_{+} &=& \frac{i}{2} \left( \mu + \frac{1}{\mu} \right)    \label{up} \\
u_{-} &=& \frac{i}{2} \left( \mu - \frac{1}{\mu} \right)    \label{um}
\end{eqnarray}
and,
\beq
\mu=\frac{k'}{k}=\sqrt{1- \frac{V}{E}}
\label{meu}
\eeq
The potential has three real parameters , $\{V_{1}, V_{2} \} \in R$ and $b \in R^{+}$. 
\paragraph{}
Fig \ref{fig_ss_complex_barrier} shows SS for a complex barrier potential $V=6.15055+35.000000045i$ of width $b=10$.  $E_{ss}$ for this potential occurs around $1256.158448057$. The figure shows that near to $E_{ss}$, $R$ and $T$ nearly overlap and at $E_{ss}$, both $R$ and $T$ diverges. We wish to analytically investigate the behavior of $R$ and $T$ around the vicinity of $E_{ss}$. The next section is dedicated for the same where we consider our complex barrier as pure imaginary for the reason of simplicity. It will be apparent in the next section that the analytical calculations are extremely lengthy and complicate even for this simple case. We have to rely on numerical investigations at those places where it is not possible to further  advance due to the transcendental nature of the equations involved. 

\section{Scattering amplitude in the neighborhood of singularity}
\label{ss_near}
We consider the simple case of complex barrier potential of imaginary height $iV$ and confined over the interval $(0,b)$  where $V \in R^{+}$. The positive value of imaginary part is an important condition to support SS \cite{ali_09}. As discussed in the previous section, SS occur when $m_{22}(E,V)=0$ ($\{E, V\} \in R^{+}$). We denote $M=\vert m_{22} \vert ^{2} $. For the pure imaginary barrier $iV$, the expression for $M$ is ,
\begin{equation}
M=Q_{1}+Q_{2}+Q_{3}+Q_{4}
\label{m_q}
\end{equation}
where,
\begin{equation}
Q_{1}=\frac{1}{2} (y_{+}^{2}+ \cos^{2}{\theta}) \cosh{2 \alpha} 
\label{q1}
\end{equation}
\begin{equation}
Q_{2}=-\frac{1}{2} (y_{-}^{2}- \sin^{2}{\theta}) \cos{2 \beta} 
\label{q2}
\end{equation}
\begin{equation}
Q_{3}=y_{-} \sin{2 \beta} \sin{\theta} 
\label{q3}
\end{equation}
\begin{equation}
Q_{4}=- y_{+} \sinh{2 \alpha} \cos{\theta}
\label{q4}
\end{equation}
In the above
\begin{equation}
y_{\pm}=\frac{k^2 \pm \rho^2}{2k \rho}
\label{yplus}
\end{equation}
\beq
\rho= (E^{2}+V^{2})^{\frac{1}{4}} \ , \theta=\frac{1}{2} \tan^{-1}{\left(\frac{V}{E} \right)}
\label{rho_theta}
\eeq
\beq
\alpha= b \rho \sin{\theta} \ , \beta=b \rho \cos{\theta}
\label{alpha_beta}
\eeq
The amplitudes of scattering 
\beq
T= \frac{1}{ M} \ , R= \frac{P}{M}
\label{tr_eq}
\eeq
where
\beq
P=\left(y_{+}^{2}-\cos ^2 {\theta }\right) \left(\sinh ^2{\alpha } \cos ^2{\beta }+\cosh ^2{\alpha } \sin ^2{\beta }\right)
\label{p_eq}
\eeq
In the next, we study the behavior of $T$ and $R$  in the neighborhood of spectral singularity. 
\subsection {Transmission amplitude in the neighborhood of SS}
\begin{figure}
\begin{center}
\includegraphics[scale=0.15]{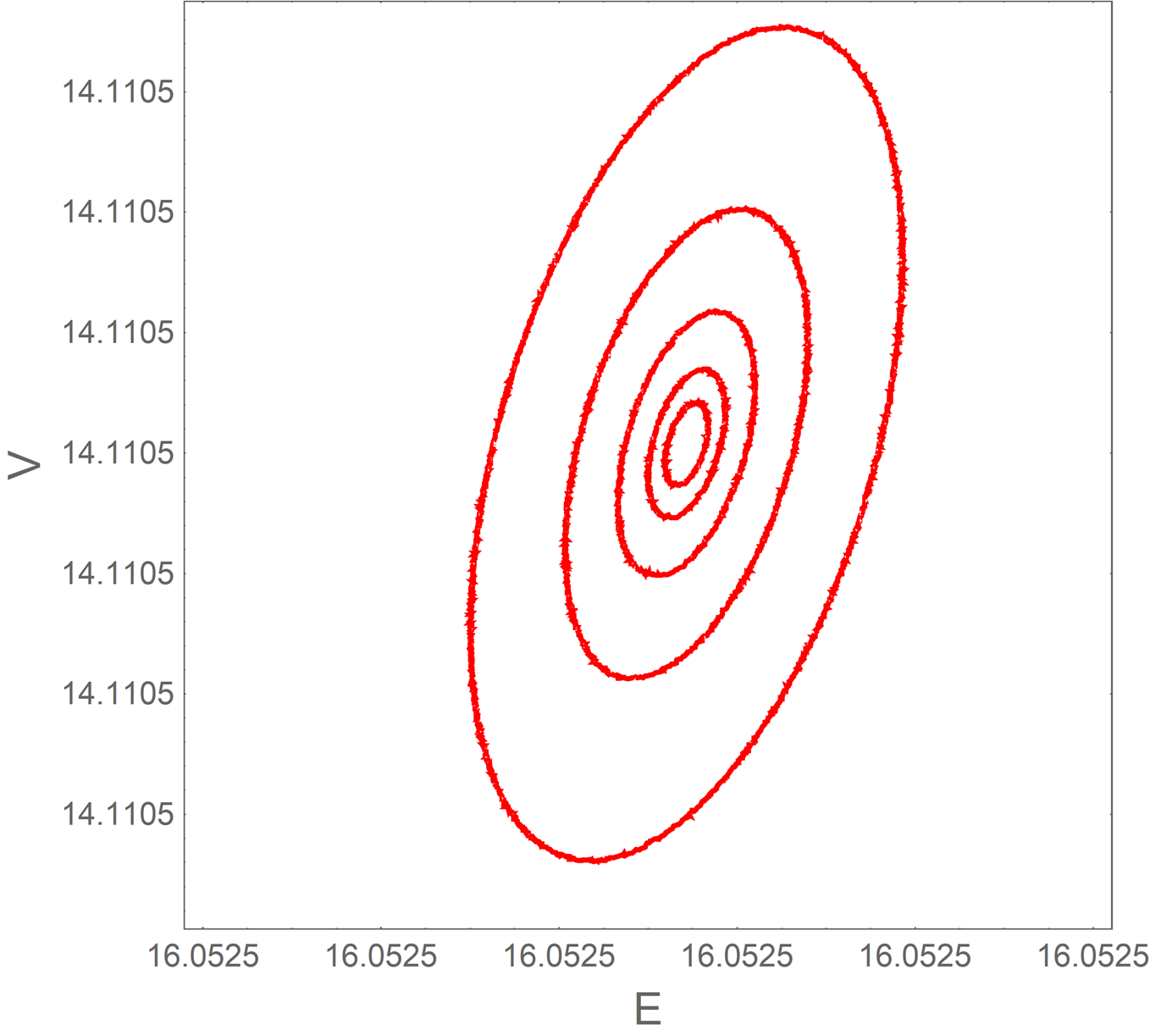}\ a  \includegraphics[scale=0.15]{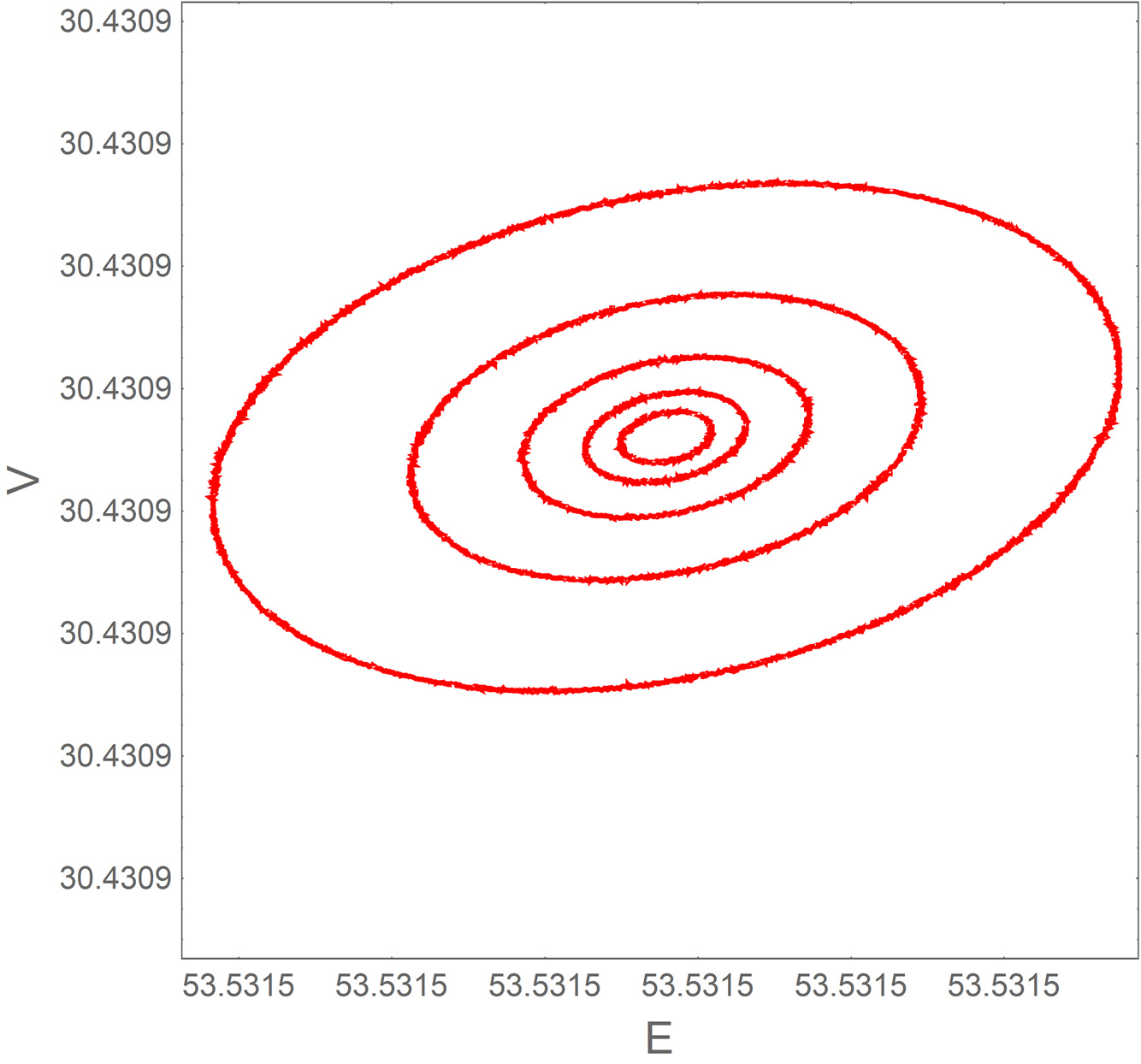}\ b 
\caption{\it The contour plot of $\log{T}$ for pure imaginary barrier of unit width around the spectral singular point $(E_{ss}= 16.052461577163, V_{ss} =14.1104958749715)$ (Fig a)and $(E_{ss}= 53.531490674672, V_{ss} =30.430864331205)$ (Fig b) in $E-V$ plane. From inward to outward, the 5 contours belongs to $\log{T}$ values of $28, 27.5, 27, 26.5$ and $26$ respectively. All these contours are concentric ellipse and self-similar to each other.  } 
\label{fig12_ss_contour}
\end{center}
\end{figure}

To understand the behavior of transmission amplitude $T=\frac{1}{M(E,V)}$ near the spectral singularity, we first investigate the behavior of quantity $M$  near the spectral singular point.  We Taylor expand $M(E,V)$ near the singular point $(E_{ss}, V_{ss})$ up to second order. The expansion  up to second order is needed to study whether the curves of constant $M$ (or constant $T$) would be closed or open in the neighborhood of $(E_{ss}, V_{ss})$. This is so because a first degree polynomial in $u=E-E_{ss}$ and  $w=V-V_{ss} $ will represent a straight line in the coordinate with $u, w$ as axis. The Taylor expansion  of $M(E,V)$ up to second order near  $(E_{ss}, V_{ss})$ gives the following equation in $u$ and $w$,
\beq
A u^{2}+B w^{2}+ 2 H u w +C u+D w+G=0
\label{conic_eq}
\eeq
where,
\beq 
A= \frac{1}{2}  \left( \frac{\partial^2  M}{\partial E^2} \right)_{E_{ss}, V_{ss}} \  ,  B= \frac{1}{2} \left (\frac{\partial^2  M}{\partial V^2} \right)_{E_{ss}, V_{ss}}  \  , H= \left( \frac{\partial^2  M}{\partial E \partial V} \right)_{E_{ss}, V_{ss}}
\eeq
\beq
 C=\left( \frac{\partial  M}{\partial E} \right)_{E_{ss}, V_{ss}} \  , D=\left( \frac{\partial  M}{\partial V} \right)_{E_{ss}, V_{ss}} \  , G=- M(E,V)
\eeq
In eq. \ref{conic_eq}, we have already used $M(E_{ss}, V_{ss})=0$. For constant-$M$ contour , the quantity $G$ is constant and Eq. \ref{conic_eq} represent equation of a conic. The nature of the conic (whether close or not) is decided by the sign of $\chi=H^2-A B$,
\beq
\chi =\left [\left( \frac{\partial^2  M}{\partial E \partial V} \right)_{E_{ss}, V_{ss}} \right]^{2} - \left [ \left( \frac{\partial^2  M}{\partial E^2} \right)_{E_{ss}, V_{ss}} \right]. \left [ \left (\frac{\partial^2  M}{\partial V^2} \right)_{E_{ss}, V_{ss}} \right]
\label{chi_eq11}
\eeq
For $\chi <0$, the conic represent a close curve. Towards this we evaluate the various derivatives of Eq. \ref{chi_eq1} below. We can express $\frac{\partial^2  M}{\partial E^2}$ as
\beq
\frac{\partial^2  M}{\partial E^2}=\frac{\partial^2  Q_{1}}{\partial E^2}+\frac{\partial^2  Q_{2}}{\partial E^2}+\frac{\partial^2  Q_{3}}{\partial E^2}+\frac{\partial^2  Q_{4}}{\partial E^2}
\label{del_m_1}
\eeq
where after lengthy algebra, we obtain
\begin{multline}
\frac{\partial^2  Q_{1}}{\partial E^2}= \frac{\cosh{2\alpha}}{8k^{6} \rho^{10}} \left( V^4(3k^4+2 \rho^4)-2k^6\rho^2V^2\cos{2\theta}-4k^8 \rho^2 V \sin{2\theta} \right)   + \\ 
\frac{b}{16 k^2 \rho^9} \left( (k^2+\rho^2)^2 +4k^2 \rho^2 \cos^2{\theta}\right) \left[ 2 g^2 b \rho \cosh{2 \alpha} + \sinh{2\alpha} (2k^2 V \cos{\theta} -(E^2-V^2)\sin{\theta})\right] \\ 
+ \frac{b \sinh{2\alpha}}{4 k^4 \rho^9} g V\left ( V^3 -2 k^4 \rho^2  \sin{2 \theta} \right)
\label{del_q1_del_e2}
\end{multline}
\begin{multline}
\frac{\partial^2  Q_{2}}{\partial E^2}= \frac{b}{16k^2 \rho^9} \left( (k^2-\rho^2)^2 -4k^2 \rho^2 \sin^2{\theta}\right) \left [ 2h^2 b \rho \cos{2\beta} -\sin{2 \beta} \left( (E^2-V^2)\cos{\theta}+ 2E V \sin{\theta} \right) \right] \\ -\frac{V \cos{2 \beta}}{8 k^6 \rho^{10}} \left [ V^3 (3 k^4+ 2 \rho^4) -2 k^6 \rho^2 V \cos{2 \theta} -4 k^8 \rho^2 \sin{2 \theta} \right] - \\  \frac{b h}{4 k^4 \rho^9} V \sin{2 \beta} (V^3-2 k^4 \rho^2 \sin{2 \theta}) 
\label{del_q2_del_e2}
\end{multline}
\begin{multline}
\frac{\partial^2  Q_{3}}{\partial E^2}= \frac{\rho^2-k^2}{8 k^5 \rho^9} \Big[ 4 b^2 k^8 \rho^2 \cos^2{\theta} \sin{\theta} \sin{2 \beta}  - 2k^2 V \sin{2 \beta} \cos{\theta} (3k^4 +2 k^2 \rho^2 +\rho^4 - 4 b^2 k^4 \rho^2 \sin^2{\theta}) \\ +\sin{2\beta} \sin{\theta} \left \{ 4b^2 k^4 \rho^2 V^2 \sin^2{\theta} -5 k^8 -8 k^6 \rho^2 -4 k^2 \rho^6 -3 \rho^8 +k^4 (V^2-4 \rho^4) \right \} \\  + b k^2 \rho \cos{2 \beta} \left \{ 4 k^4 V \cos^2{\theta} +4V \sin^2{\theta} (2 k^4+2 k^2 \rho^2 +\rho^4) +k^2 \sin{2\theta} (5 k^4+ 4k^2 \rho^2 +3 V^2) \right \} \Big ]  
\label{del_q3_del_e2}
\end{multline}
\begin{multline}
\frac{\partial^2  Q_{4}}{\partial E^2}= -\frac{(\rho^2+k^2)}{8 k^5 \rho^9} \Big[ 4 bk^2 \rho \cosh{2 \alpha} \big \{ (2k^4-2k^2 \rho^2 +\rho^4) V \cos^2{\theta} +k^4 V \sin^2{\theta} + \\ \frac{k^4 \sin{2 \theta}}{4} (4 k^2 \rho^2 -5k^4-3V^2)  \big \} + \cos{\theta} \sinh{2 \alpha} \big \{ (k^2-\rho^2)^2 (5 k^4+2 k^2 \rho^2 +3 \rho^4)-k^4 V^2 \big \}  \\ + 4 b^2 k^4 \rho^2 V^2 \sinh{2 \alpha} \cos^3{\theta} + 4 b^2 k^8 \rho^2 \cos{\theta} \sin^2{\theta} \sinh{2 \alpha} \\ -2 k^2 V \sin{\theta} \sinh{2 \alpha}(3 k^4 -2k^2 \rho^2 +\rho^4 + 4 b^2 k^4 \rho^2 \cos^2 {\theta})      \Big ]  
\label{del_q4_del_e2}
\end{multline}
In Eqs. \ref{del_q1_del_e2}, \ref{del_q2_del_e2},  \ref{del_q3_del_e2} and  \ref{del_q4_del_e2} we have used the following notations
\beq
g=V \cos{\theta}-E \sin{\theta}
\label{g_expression}
\eeq
\beq
h=V \sin{\theta}+E \cos{\theta}
\label{h_expression}
\eeq
Using Eqs. \ref{del_q1_del_e2}, \ref{del_q2_del_e2},  \ref{del_q3_del_e2} and  \ref{del_q4_del_e2} in the Eq. \ref{del_m_1},  $\frac{\partial^2  M}{\partial E^2}$ can be calculated. The expressions for  $\frac{\partial^2  M}{\partial V^2}$  is given by,
\begin{multline}
\frac{\partial^2  M}{\partial V^2}= \\
\frac{1}{8 \rho^8}\Big[ 2\sin{2\beta} \sin{\theta} \{ V^2(y_{-}+4 y_{+})-2 \rho^4 y_{+}\} -4 V y_{+} (E \cos{\theta}\sin{2\beta} +2 b \rho \cos{2\beta} \sin{\theta} g) \\ 
-b \rho (1-2 y_{-}^2 -\cos{2\theta}) \{ 2 b \rho g^2 \cos{2\beta}+ \sin{2\beta} ( (E^2-V^2)\cos{\theta} +2 E V \sin{\theta}) \} \\ + 8 y_{-} b E \rho \cos{2\beta} \cos{\theta} g -2y_{-} E \sin{2\beta} (3 V +f) \\
 -4 y_{-} b \rho \sin{\theta} \{ 2 g^2 b \rho \sin{2\beta} -\cos{2\beta} ((E^2-V^2)\cos{\theta} +2 E V \sin{\theta})\} \\  - 2(\cosh{2\alpha}-\cos{2\beta}) \Big( \frac{3 }{2 \sqrt{2}} E V^2 -E^2 \cos{2\theta} +2 E V \sin{2\theta}\Big) \\ 
+2 E y_{+} \sinh{2\alpha}(p-3 V \sin{\theta}) -4 b \rho h \sinh{2\alpha} (2 V y_{-} y_{+} +E \sin{2\theta} ) \\ 
+ 4 E \sin{\theta} (2 b \rho y_{+} h \cosh{2\alpha} -V y_{-} \sinh{2\alpha}) -4 b \rho \sin{2\beta} g(2 V y_{-} y_{+} + E \sin{2\theta}) \\
+ 2 b\rho (y_{+}^2 +\cos^2{\theta}) \{ 2 h^2 b \rho \cosh{2\alpha} -\sinh{2\alpha} (2 E V \cos{\theta} -(E^2-V^2) \sin^2{\theta})\} \\ 
+8 b h \rho y_{-} V \cos{\theta} \cosh{2\alpha} +2 \cos{\theta} \sinh{2\alpha} (2 r^4 y_{-} -V^2 (4 y_{-} +y_{+})) \\ - 4 b \rho y_{+}\cos{\theta} \{ 2 b\rho h^2 \sinh{2\alpha} -\cosh{2\alpha} (2 E V \cos{\theta} -(E^2-V^2)\sin{\theta}) \}  \Big]
\label{del_m_del_v2}
\end{multline}
Where $f$ and $p$ are given by,  
\beq
f=V \cos{\theta}+ E \sin{\theta}
\label{f_expression}
\eeq
\beq
p=E \cos{\theta}- V \sin{\theta}
\label{f_expression}
\eeq
After a lengthy algebra we obtain the expression of $\frac{\partial^{2} M}{ \partial E \partial V}$ and is given below in terms of the functions $g_{i}, i \in \{1,2,3,4,5\}$,
\beq
\frac{\partial^{2} M}{ \partial E \partial V} = \frac{1}{16 k^{4} \rho^{10}} \left [(2 g_{1}+ k \rho^{2}g_{2}) \cos{2 \beta} + g_{3} \cosh{2 \alpha} + \rho (g_{4} \sin{2 \beta} +g_{5} \sinh{2 \alpha} ) \right]
\label{d2mdedv}
\eeq
where,
\begin{multline}
g_{1}=2 b^2 k^6 \rho^4 V \cos ^2{2 \theta }+k^3 \rho^2 V \cos {2 \theta } \left[2 b^2 k^3 \left(k^2-2 \rho^2\right)+b^2 k V^2+2 b V \left(k^2+\rho^2\right)-2 k^3\right] \\ -2 b k^5 \rho^2 \left(-k^4+k^2 \rho^2+2 V^2\right)+V^3 \left(3 k^4+\rho^4\right)
\end{multline}
\begin{multline}
g_{2}=b^{2} k^{3} \rho^{2} \sin {4 \theta } \left( V^{2}-k^{4}\right ) - \sin {2 \theta } [ b^{2} k ( k^{4}-4 k^{2} \rho^{2}+\rho^{4} ) ( k^{4}-V^{2}  ) \\ +  2 b V ( k^{2}+\rho^{2} ) (2 k^{4}-\rho^{4} ) +4 k^{3} ( \rho^{4}-2 V^{2} ) ]
\end{multline}
\begin{multline}
g_{3} = -2 b^2 k^6 \rho^4 V+6 k^8 V-4 k^4 \rho^4 V-2 \rho^8 V + 2 b k^3 \rho^2 [k^6-k^4 \rho^2+k^2 (\rho^4-5 V^2)+3 (\rho^6-\rho^2 V^2)] \\
+k \rho^{2} \Big[ \sin{2 \theta} \big \{ b^2 k \left(k^4+4 k^2 \rho^2+\rho^4\right) \left(k^4-V^2\right)+2 b V (k^2-\rho^2)  \left(2 k^4-\rho^4\right)+4 k^3 \left(\rho^4-2 V^2\right)  \big \} \\
- 2 k^2 \cos{2 \theta } \big  \{ b^2 k V \left(k^4+4 k^2 \rho^2+\rho^4\right)+b \left(k^6+3 k^4 \rho^2-k^2 \rho^4+V^2 \left(3 k^2+\rho^2\right)-3 \rho^6\right)-2 k^3 V  \big \} \\ 
 + b^2 k^3 \rho^2 \sin {4 \theta } (k^4-V^2) -2 b^2 k^5 \rho^2 V \cos {4 \theta }  \Big]
\end{multline}
\begin{multline}
g_{4}= 2 b k^3 \rho^2 \cos {3 \theta } \left[k^3 V-b (k^2-\rho^2)  (k^4-V^2 )\right] + \\  2 \cos{\theta} \Big [ k^{3} (k^2 -\rho^{2})\{-2 k^2 \rho^2 - 3 \rho^4 + k^4 (-1 + b^2 \rho^2)\} + b V  (   -3 k^8+k^6 \rho^2+2 k^4 \rho^4-\rho^8 )  + \\ k^3 V^2 \{ -\rho^2 (b^2 k^2+3)+b^2 \rho^4+5 k^2 \}  + \Big] \\ 
 k \sin{\theta}\Big [ 2 V (6 k^6+2 k^4 \rho^2-k^2 \rho^4+\rho^6)  + b k \{k^8-3 k^4 \rho^4-8 k^2 \rho^6-V^2 \left(5 k^4-16 k^2 \rho^2+\rho^4\right)+4 \rho^8\} \\  -2 b k^3 \rho^2 \cos {2 \theta } \{4 b k V (k^2-\rho^2)+3 k^4-2 \rho^4+V^2 \} \Big]
\end{multline}
\begin{multline}
g_{5}=  -  2 \sin{\theta} \Big [ k^3 \left(k^2+\rho^2\right) \{k^2 \rho^2 \left(b^2 k^2-2\right)+k^4+3 \rho^4\} + b V \left(3 k^8+k^6 \rho^2-2 k^4 \rho^4+\rho^8\right) \\ -k^3 V^2 \{\rho^2 \left(b^2 k^2+3\right)+b^2 \rho^4+5 k^2\}   \Big]   \\  +  k \cos{\theta} \Big [ -b k \left(k^8-3 k^6 \rho^2-3 k^4 \rho^4+10 k^2 \rho^6+4 \rho^8\right) + b k V^2 \left(5 k^4+17 k^2 \rho^2+\rho^4\right) \\ + 2 V \{2 k^6 \left(b^2 \rho^2-3\right)+2 k^4 \left(b^2 \rho^4+\rho^2\right)+k^2 \rho^4+\rho^6 \}  \Big] \\ + b k^4 \rho^2 \cos {3 \theta } \left [4 b k V \left(k^2+\rho^2\right)+3 k^4-2 \rho^4+V^2 \right ] \\  + 2 b k^3 \rho^2 \sin {3 \theta } \left[k^3 V-b \left(k^2+\rho^2\right) \left(k^4-V^2\right)\right]
\end{multline}
To find the sign of $\chi$, we have to evaluate the right hand side of Eq. \ref{chi_eq11} at the spectral singular points $(E_{ss}, V_{ss})$. Analytically, this is expected to be an extremely non-trivial calculation  due to the transcendental relations between $E_{ss}$ and $V_{ss}$ as well as the complicate expressions for $\chi$. Therefore in the current work we  calculate $\chi (E_{ss}, V_{ss})$ numerically to find the sign of $\chi$. In Table \ref{table1} we list ten spectral singularities $(E_{ss}, V_{ss})$ for pure imaginary potential of width $b=1$ and calculate the values of $\chi (E_{ss}, V_{ss})$. Highly accurate values of $(E_{ss}, V_{ss})$ are provided in the table to calculate $\chi$. It can be seen from Table \ref{table1} that for all cases $\chi(E_{ss}, V_{ss})<0$. It can be also be shown that for these cases $A(E_{ss}, V_{ss}) \neq B(E_{ss}, V_{ss})$ (see Table \ref{table2}). Thus the constant-$M$ contour and therefore (constant- $T$ contours ) in the neighborhood of spectral singularities $(E_{ss}, V_{ss})$ are ellipse. The center of all these concentric ellipse for different values of $M$ in the neighborhood of  $(E_{ss}, V_{ss})$ have the same eccentricity as the conic equation \ref{conic_eq} only differ in $G$ for different (constant) $M$. Therefore all such elliptical contours are self-similar.  
\begin{table}
\begin{center}
\caption{ Value of $\chi$ at spectral singular point for pure imaginary barrier of width $b=1$}
\begin{tabular} {|c|c|c|c|}
\hline 
 SS &  \textbf{$E_{ss}$} & \textbf{$V_{ss}$} &  \textbf{$\chi( E_{ss}, V_{ss})$}  \\
\hline
\textbf{$SS_{1}$} & $16.052461577163$ & $14.1104958749715$ &  $-2.83207 \times 10^{-5}$   \\
\hline
\textbf{$SS_{2}$} & $53.531490674672$ & $30.430864331205$ &  $-7.1323 \times 10^{-7}$  \\
\hline
\textbf{$SS_{3}$} & $111.265616666905$ & $48.524980456215$ &  $-6.63487 \times 10^{-8}$  \\
\hline
\textbf{$SS_{4}$} & $189.01989429134$ & $67.933020370195$ & $-1.1295 \times 10^{-8}$  \\
\hline
\textbf{$SS_{5}$} & $286.69487833905$ & $88.38604799591$ & $-2.7397 \times 10^{-9}$  \\
\hline
\textbf{$SS_{6}$} & $404.23765595182$ & $109.70756999454$ & $-8.40342 \times 10^{-10}$  \\
\hline
\textbf{$SS_{7}$} & $541.6162666414$ & $131.7726934002$ & $-3.04622 \times 10^{-10}$  \\
\hline
\textbf{$SS_{8}$} & $698.80974645375$ & $154.4882891468$ & $-1.25157 \times 10^{-10}$  \\
\hline
\textbf{$SS_{9}$} & $875.8035169372$ & $177.78219065599$ & $ -5.66853 \times 10^{-11} $  \\
\hline
\textbf{$SS_{10}$} & $1072.5869864069$ & $201.5968028915$ & $-2.77573 \times 10^{-11} $  \\
\hline
\end{tabular} 
\label{table1}
\end{center}
\end{table}
The constant-M contours around the spectral singular point are of same shape as the `corresponding' constant-T contours (due to $T=1/M$). Thus the tunnelling amplitude in the neighborhood of spectral singular point are the self-similar concentric ellipse.  In Fig \ref{fig12_ss_contour} we plot the contours for $\log{T}$ values as $28, 27.5, 27, 26.5$ and $26$ around the spectral singular points $(E_{ss}= 16.052461577163, V_{ss} =14.1104958749715)$ (Fig \ref{fig12_ss_contour}-a)and $(E_{ss}= 53.531490674672, V_{ss} =30.430864331205)$ (Fig \ref{fig12_ss_contour}-b). It is seen that all the contours are concentric self -similar ellipse.  
\subsection{Reflection amplitude in the neighborhood of SS}
We write the reflection amplitude as
\beq
R=\frac{1}{S(E,V)}
\label{r_s_eq}
\eeq
where
\beq
S=\frac{M}{P}
\label{s_eq}
\eeq
\begin{figure}
\begin{center}
\includegraphics[scale=0.45]{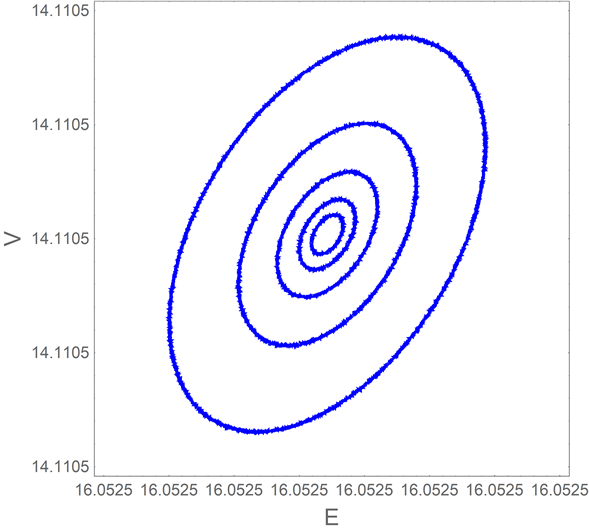}\ a  \includegraphics[scale=0.45]{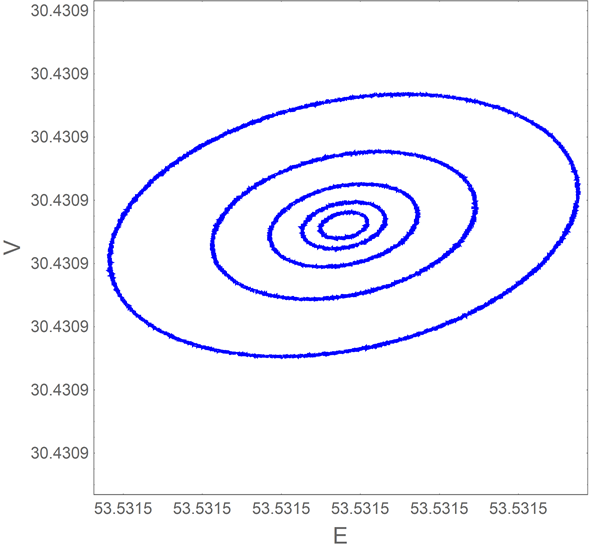}\ b 
\caption{\it The contour plot of $\log{R}$ for pure imaginary barrier potential of unit width around the spectral singular point $(E_{ss}= 16.052461577163, V_{ss} =14.1104958749715)$ (Fig a)and $(E_{ss}= 53.531490674672, V_{ss} =30.430864331205)$ (Fig b) in $E-V$ plane. From inward to outward, the 5 contours belongs to $\log{R}$ values of $28, 27.5, 27, 26.5$ and $26$ respectively. All these contours are concentric ellipse and self-similar to each other.  } 
\label{fig_r_ss_contour}
\end{center}
\end{figure} 
It can be noted that at SS, $S(E_{ss}, V_{ss})=0$. To investigate the behavior of $R$ near SS, we Taylor expand `$S$' near SS and using  $S(E_{ss}, V_{ss})=0$ we obtain the second degree polynomial in $u$ and $w$ as,  
\beq
A' u^{2}+B' w^{2}+ 2 H' u w +C' u+D' w+G'=0
\label{conic_eq_r}
\eeq
where,
\beq 
A'= \frac{1}{2}  \left( \frac{\partial^2  S}{\partial E^2} \right)_{E_{ss}, V_{ss}} \  ,  B'= \frac{1}{2} \left (\frac{\partial^2  S}{\partial V^2} \right)_{E_{ss}, V_{ss}}  \  , H'= \left( \frac{\partial^2  S}{\partial E \partial V} \right)_{E_{ss}, V_{ss}}
\eeq
\beq
 C'=\left( \frac{\partial  S}{\partial E} \right)_{E_{ss}, V_{ss}} \  , D'=\left( \frac{\partial  M}{\partial V} \right)_{E_{ss}, V_{ss}} \  , G'=- S(E,V)
\eeq
For constant amplitude $R$, the quantity $S$ is also constant and therefore eq. \ref{conic_eq_r} represent a conic. Again the nature of these conics for  different fixed value of $S$ are decided by the sign of   
\beq
\chi'= H'^{2}- A' B' =\left [\left( \frac{\partial^2  S}{\partial E \partial V} \right)_{E_{ss}, V_{ss}} \right]^{2} - \left [ \left( \frac{\partial^2  S}{\partial E^2} \right)_{E_{ss}, V_{ss}} \right]. \left [ \left (\frac{\partial^2  S}{\partial V^2} \right)_{E_{ss}, V_{ss}} \right]
\label{chi_eq}
\eeq
Next we evaluate $A', B'$ and $H'$ to investigate the sign of $\chi '$. These expressions are provided below
\beq
H' = \frac{H}{P}- \frac{1}{P^{2}} (C G_{2}+D G_{1})  
\label{hprime_ess}
\eeq  
\beq
A'= - \frac{C G_{1}}{P^{2}} +\frac{A}{P}
\label{a_prime}  
\eeq
\beq
B'=  - \frac{D G_{2}}{P^{2}} +\frac{B}{P}
\label{b_prime}  
\eeq
Here,
\beq
G_{1}= \frac{\partial P}{ \partial E} \ , G_{2}= \frac{\partial P}{ \partial V} \
\eeq
We arrived at eqs.\ref{hprime_ess}, \ref{a_prime} and \ref{b_prime} by directly expanding the derivatives of $S$ in terms of the derivatives of $M$ and $P$ and using $M (E_{ss}, V_{ss})=0$.   All the quantities in eq. \ref{hprime_ess}, \ref{a_prime} and  \ref{b_prime} are to be evaluated at $(E_{ss}, V_{ss})$. Expressions for $G_{1}$, $G_{2}$ are given below
\begin{multline}
G_{1}=\frac{1}{8 k^4 \rho^{10}}  \big[ 4 b \rho^7 k^4  (y_{+}^2-\cos ^2\theta ) (h \sin {2 \beta} -g \sinh {2 \alpha }) \\ +  \rho^4 V (\cos {2 \beta }-\cosh {2 \alpha }) ( +V^3+ 2 k^4 \rho^2 \sin {2 \theta } ) \big ]
\label{g1_eq}
\end{multline}
\begin{multline}
G_{2}=\frac{1}{8 k^2 \rho^{6}}  \big[ 4 b \rho^3 k^2  (y_{+}^2-\cos ^2\theta ) (h \sin {2 \alpha} +g \sin {2 \beta }) \\ -  (  \cos {2 \beta }-\cosh {2 \alpha }) ( V^3 +2 k^4 \rho^2 \sin {2 \theta } ) \big ]
\label{g2_eq}
\end{multline}
Now we have the expressions for all the quantities required to evaluate the value of $\chi'$. We find that $\chi' (E_{ss}, V_{ss})=\chi (E_{ss}, V_{ss}) <0$ and thus the constant-$S$ (and therefore constant-$R$) contours are ellipses. The equality of $\chi=\chi'$ at SS  is extremely surprising. For the clarity we calculate the coefficients $A, B, H$ and $A',B',H'$ in Table \ref{table2}. From the table it is seen that $A=A'$, $B=B'$ and $H=H'$ and thus $\chi=\chi'$. The equality of these coefficients only hold at SS. $\chi=\chi'$ at SS  also shows that the orientations of constant-$T$ and constant-$R$ contours in the neighborhood of SS are same . Constant-$R$ contours in the neighborhood of SS are also self-similar ellipse as the conic represented by  eq. \ref{conic_eq_r} only differs from each other in the value of $S$. In Fig \ref{fig_r_ss_contour} we plot the contours for $\log{R}$ values of $28, 27.5, 27, 26.5$ and $26$ around the spectral singular points $(E_{ss}= 16.052461577163, V_{ss} =14.1104958749715)$ (Fig \ref{fig_r_ss_contour}-a)and $(E_{ss}= 53.531490674672, V_{ss} =30.430864331205)$ (Fig \ref{fig_r_ss_contour}-b). All these constant- $R$ contours are concentric self -similar ellipse. 
\begin{table}
\begin{center}
\caption{ Value of coefficients $A, A', B, B', H, H'$  at spectral singular point for pure imaginary barrier potential of width $b=1$}
\begin{tabular} {|l|l|l|l|l|l|l| }
\hline 
 SS &  \textbf{$A \times 10^{-5}$} & \textbf{$A' \times 10^{-5}$} &  \textbf{$B \times 10^{-5}$} & \textbf{$B' \times 10^{-5}$} &   \textbf{$H \times 10^{-5}$} & \textbf{$H' \times 10^{-5}$}  \\
\hline
 \textbf{$SS_{1}$} &  \textbf{$211.999 $} & \textbf{$211.999 $} &  \textbf{$417.192$} & \textbf{$417.192 $} &   \textbf{$-265.651$} & \textbf{$-265.651$}  \\
\hline
\textbf{$SS_{2}$} &  \textbf{$30.6455 $} & \textbf{$30.6455 $} &  \textbf{$62.5316 $} & \textbf{$62.5316 $} &   \textbf{$-23.0855 $} & \textbf{$-23.0855 $}  \\
 \hline
  \textbf{$SS_{3}$} &  \textbf{$9.29923 $} & \textbf{$9.29923$} &  \textbf{$18.4784 $} & \textbf{$18.4784 $} &   \textbf{$-4.88389 $} & \textbf{$-4.88389 $}  \\
  \hline
  \textbf{$SS_{4}$} &  \textbf{$3.86239 $} & \textbf{$3.86239 $} &  \textbf{$7.46395 $} & \textbf{$7.46395 $} &   \textbf{$-1.54586 $} & \textbf{$-1.54586 $}  \\
  \hline
  \textbf{$SS_{5}$} &  \textbf{$1.91855 $} & \textbf{$1.91855 $} &  \textbf{$3.61982 $} & \textbf{$3.61982 $} &   \textbf{$-0.618261 $} & \textbf{$-0.618261 $}  \\
  \hline
  \textbf{$SS_{6}$} &  \textbf{$1.07116 $} & \textbf{$1.07116 $} &  \textbf{$1.9807 $} & \textbf{$1.9807 $} &   \textbf{$-0.288363 $} & \textbf{$-0.288363 $}  \\
  \hline
 \textbf{$SS_{7}$} &  \textbf{$0.649646 $} & \textbf{$0.649646 $} &  \textbf{$1.18092 $} & \textbf{$1.18092 $} &   \textbf{$-0.149991 $} & \textbf{$-0.149991 $}  \\
  \hline
  \textbf{$SS_{8}$} &  \textbf{$0.419131 $} & \textbf{$0.419131$} &  \textbf{$0.750798 $} & \textbf{$0.750798 $} &   \textbf{$-0.0846404 $} & \textbf{$-0.0846404$ } \\
 \hline
 \textbf{$SS_{9}$} &  \textbf{$0.283713 $} & \textbf{$0.283713 $} &  \textbf{$0.501777 $} & \textbf{$0.501777 $} &   \textbf{$-0.0508817 $} & \textbf{$-0.0508817 $ } \\
  \hline
  \textbf{$SS_{10}$} &  \textbf{$0.199567 $} & \textbf{$0.199567 $} &  \textbf{$0.349016 $} & \textbf{$0.349016 $} &   \textbf{$-0.0321738 $} & \textbf{$-0.0321738 $}  \\
 \hline
\end{tabular} 
\label{table2}
\end{center}
\end{table}
In the Fig \ref{fig_tr_ss5} we plot $\log{T}=25=\log{R}$ around the spectral singular point $SS_{5}$ of Table \ref{table1}. Both the curves overlap each others and thus verify the same orientations of the constant-$T,R$ contours around SS. We have verified this at other values of SS for the present problem.
\begin{figure}
\begin{center}
\includegraphics[scale=0.25]{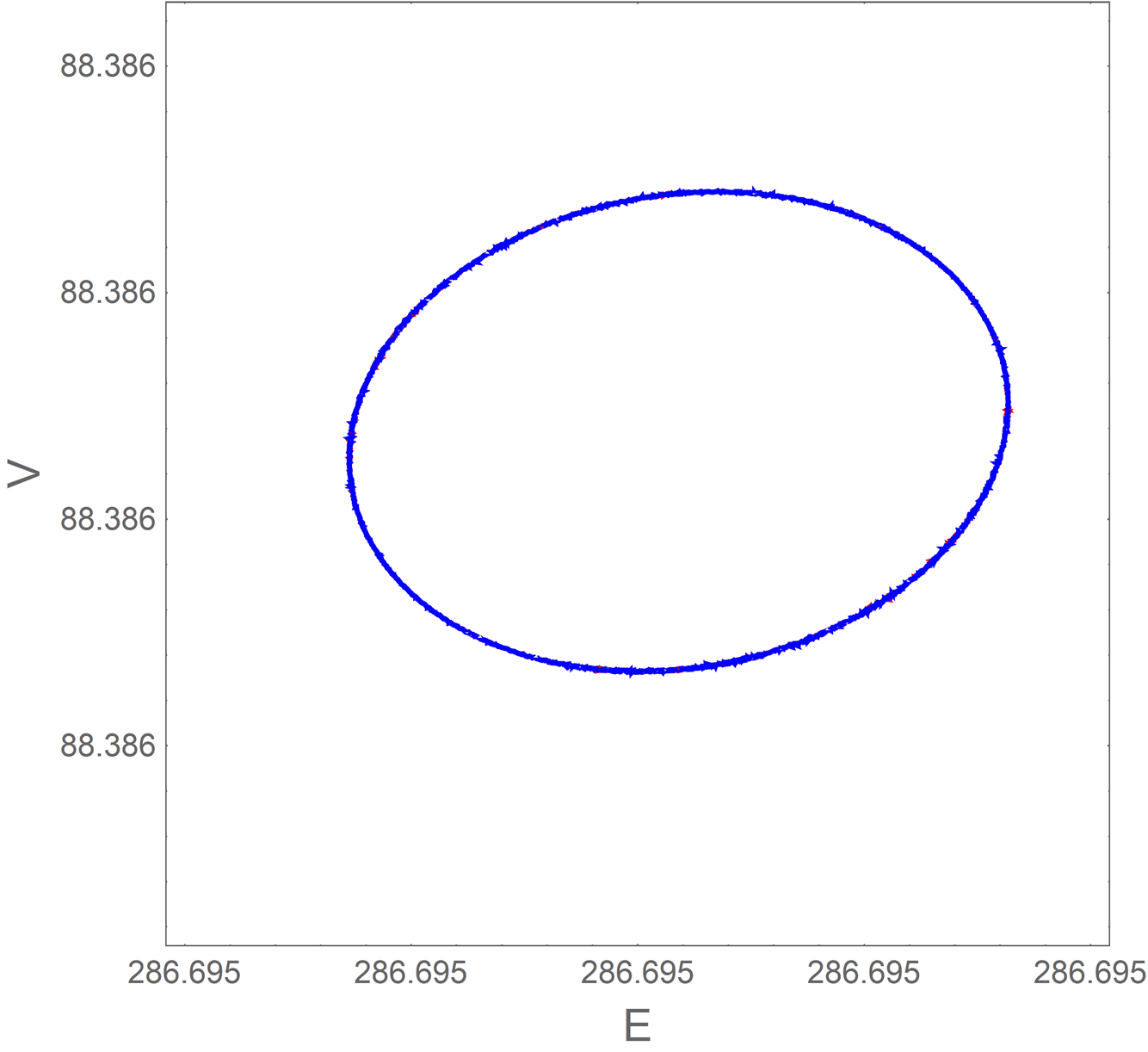} 
\caption{\it The contour plot of $\log{T}=25$ and $\log {R}=25$ for pure imaginary barrier potential of unit width around the spectral singular point $(E_{ss}=286.69487833905 , V_{ss}=88.38604799591$  in $E-V$ plane. Both curves overlap on each other } 
\label{fig_tr_ss5}
\end{center}
\end{figure} 
\begin{figure}
\begin{center}
\includegraphics[scale=0.25]{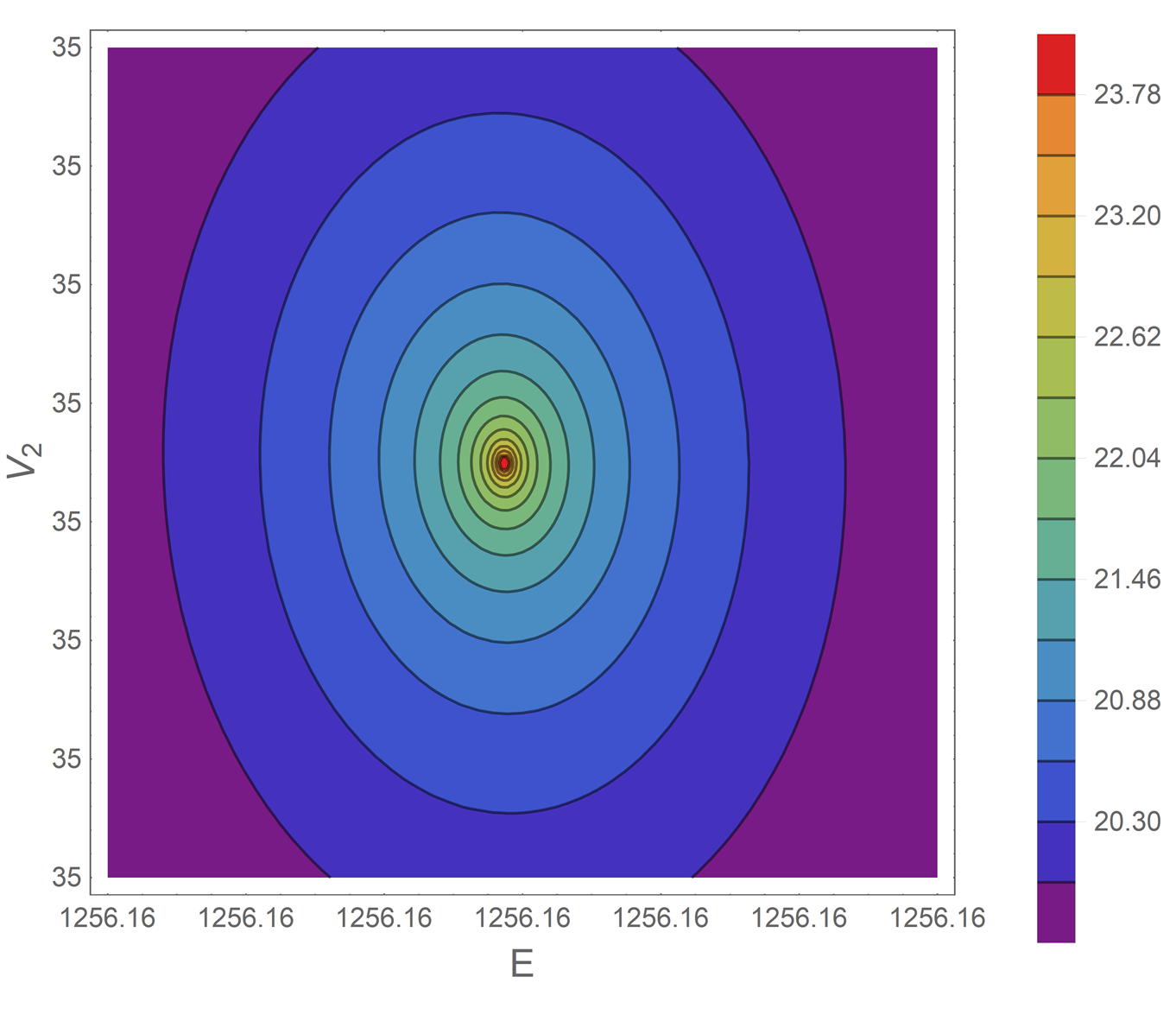} 
\caption{\it The contour plot of $\log{T}$ for the potential $V=6.15055+V_{2}i$ of width $b=10$ near the spectral singular point. The contours of constant $T$ (shown here in $E-V_{2}$ plane) are close curve around spectral singularity and appears to be self-similar. } 
\label{fig_ss_contour}
\end{center}
\end{figure} 
In the present study, we have shown the behavior of the loci of constant-$T$ and $R$ in the  infinitesimal vicinity of SS for the complex potential of the form $iV$, $V \in R^{+}$. The calculations of such loci are extremely complicate for barrier potential of the form $V= V_{1}+ i V_{2}$ and therefore we have left it in the present study. However, the barrier potential of this form also appears to have self-similar elliptical type behavior for constant- $T$ and $R$ contours around SS. We demonstrate this graphically in Fig \ref{fig_ss_contour} for the potential $V=6.15055+ iV_{2}$ of  width $b=10.0$ around the spectral singularity $E_{ss}= 1256.158448057, V_{2, SS}=35.000000045$ (same potential as of Fig \ref{fig_ss_complex_barrier}). All the constant- $T$ contours in this figure are self-similar ellipse. 
\section{Results and Discussions}
\label{result_discussion}
There have been extensive study in the spectral singularity in the last one and half decade. However the behavior of the scattering amplitude in the vicinity of the spectral singularity have not been studied before. Such studies are important for the practical applications of spectral singularity (SS) and coherent perfect absorption (CPA) as the system parameters are extremely difficult to be maintained around those values that corresponds to SS or CPA.  We have studied the behavior of the transmission and reflection amplitude in the vicinity of the spectral singularity for a pure imaginary barrier. By Taylor expansion of the quantities $M$ and $S$ ($M=\frac{1}{T}, S= \frac{1}{R}$) up to second order ($T, R$ are transmission and reflection  amplitude), we have derived the equation of the conic for constant-$M$ and constant-$S$ contours in the neighborhood of the spectral singularity for a pure imaginary barrier.  It is shown that these conic (or loci)  are close elliptical curves . It is further shown that all such ellipse are concentric and have the same eccentricity and thus are self-similar. 

The orientations of all elliptical contours for $R$ and $T$ are also same which is an extremely surprising result due to the different mathematical forms of $R$ and $T$. The contours of $R$ and $T$ are found to overlap in the vicinity of SS.  Therefore the scattering amplitude in the immediate vicinity of the spectral singularity behaves in a self-similar fashion for pure imaginary barrier potential. We expect this result would be true for general complex barrier potential $V=V_{1}+i V_{2}$. However due to mathematical complexity involved in deriving the conic equation in the vicinity of SS, we have not studied it presently. The graphical representation show such elliptical contours for a general complex barrier potential.  It is to be noted that the self-similar behavior of $T$ and $R$ in the vicinity of SS is not present for the case of delta potential (we left this exercise to interested reader) and presently it is not known what class of potential shows self-similar features of scattering amplitude in the vicinity of SS.    
  
{ \bf Acknowledgments}: 

MH acknowledges support from Dr Shyama Narendranath KC from SAG-URSC/ISRO and Director-SSPO/ISRO HQ for their supports.  BPM acknowledges the support from MATRIX project (Grant No. MTR/2018/000611), SERB, DST
Govt. of India.

\end{document}